\documentclass{JHEP3}
\usepackage{units}
\usepackage{graphicx,epsfig}
\usepackage{amsmath}
\newcommand{\Imag}{\mathop\mathrm{Im}}

\newcommand{\keV}{\mathrm{~keV}}
\newcommand{\MeV}{\mathrm{~MeV}}
\newcommand{\GeV}{\mathrm{~GeV}}

\newcommand{\be}{\begin{equation}}
\newcommand{\ee}{\end{equation}}
\newcommand{\ba}{\begin{eqnarray}}
\newcommand{\ea}{\end{eqnarray}}
\renewcommand{\d}{\partial}
\renewcommand{\l}{\left(}
\renewcommand{\r}{\right)}
\newcommand{\la}{\langle}
\newcommand{\ra}{\rangle}
\newcommand{\e}{\mathrm{e}}
\newcommand{\Br}{\mathop\mathrm{Br}}
\renewcommand{\L}{\mathcal{L}}
\newcommand{\half}{\frac{1}{2}}

\title{Light inflaton Hunter's Guide}

\author{F. Bezrukov\\
  Max-Planck-Institut f\"ur Kernphysik,\\
  PO Box 103980, 69029 Heidelberg, Germany\\
  Institute for Nuclear Research of the Russian Academy of Sciences,\\
  60th October Anniversary prospect 7a, Moscow 117312, Russia\\
  E-mail: \email{Fedor.Bezrukov@mpi-hd.mpg.de}}
\author{D. Gorbunov\\
  Institute for Nuclear Research of the Russian Academy of Sciences,\\
  60th October Anniversary prospect 7a, Moscow 117312, Russia\\
  E-mail: \email{gorby@ms2.inr.ac.ru}
}

\date{4 May 2010}

\abstract{
  We study the phenomenology of a realistic version of the
  chaotic inflationary model, which can be fully and directly explored
  in particle physics experiments. The inflaton mixes with the
  Standard Model Higgs boson via the scalar potential, and no
  additional scales above the electroweak scale are present in the
  model.  The inflaton-to-Higgs coupling is responsible for both
  reheating in the Early Universe and the inflaton production in
  particle collisions.  We find the allowed range of the light inflaton
  mass, $270\MeV\lesssim m_\chi\lesssim 1.8$~GeV, and discuss the ways
  to find the inflaton.  The most promising are two-body kaon and
  $B$-meson decays with branching ratios of orders $10^{-9}$ and $
  10^{-6}$, respectively.  The inflaton is unstable with the lifetime
  $10^{-9}$--$10^{-10}$~s.  The inflaton decays can be searched for in a
  beam-target experiment, where, depending on the inflaton mass,
  from several billions to several tenths of millions 
inflatons can be produced per
  year with modern high-intensity beams.
}

  
\begin{document}

%%%%%%%%%%%%%%%%%%%%%%%%%%%%%%%%%%%%%%%%%%%%%%%%%%%%%%%%%%%%%%%%%%%%%%%%
\section{Introduction}
\label{sec:intro}

In this paper we present an example of how (low energy) particle
physics experiments can directly probe the inflaton sector (whose
dynamics is important at high energies in the very Early Universe).

The common assumption about the inflaton sector is that it is
completely decoupled from the Standard Model (SM) at energies much
lower than the inflationary scale.  This assumption appears quite
natural, since the slow-roll conditions generally permit only tiny
coupling of the inflaton to any other fields including itself (see
e.g.\ \cite{Lyth:1998xn} for a review).  This implies within the
perturbative approach that by integrating out the inflaton sector one
obtains at low energies some non-renormalisable operators strongly
suppressed by both tiny couplings and high inflationary scale.  These
assumptions prevent any direct (laboratory) investigation of the
inflationary mechanism.

The situation is quite different if the inflaton sector contains only
light fields.  In this case even weak inflaton coupling to the SM
particles can lead to observable signatures in laboratory experiments.

We will concentrate here on the model, which is a particular version
of the simple chaotic inflation with quartic potential and with
inflaton field coupled to the SM Higgs boson via a renormalisable
operator.  We confine ourselves to the case where the Higgs scalar
potential is scale-free at the tree level, so that its vacuum
expectation value is proportional to that of the inflaton field and
Higgs boson mixes with inflaton.  In this model, the inflaton $\chi$
is found to be light\footnote{In fact, there is another window for the
  inflaton mass, which is also below 1~TeV, see
  eqs.~\eqref{heavy-inflaton-upper-limit},
  \eqref{heavy-inflaton-lower-limit}.}, $270\MeV\lesssim
m_\chi\lesssim 10$~GeV, where the lower limit actually comes from the
searches for decays $K\to\pi+\text{nothing}$ and from the search for
axion-like particles in CHARM experiment, and the upper limit is
related to the requirement of a sufficient reheating after inflation.
The most promising here are searches of inflaton in decays of $K$- and
$B$-mesons and searches for inflaton decays in beam-target
experiments.

We present the estimates of meson decay branching ratios into the
inflaton, and give the estimate for the inflaton production rate in
beam-target experiment with beam parameters of the T2K, NuMi, CNGS and
NuTeV experiments \cite{T2K,NuMi,CNGS,NuTeV}.  The interesting
branchings start from $10^{-6}$ for the decay $B\to K+\chi$ and
$10^{-9}$ for the decay $K\to \pi+\chi$ (see
section~\ref{sec:mesons}).  The production rate in the beam-target
experiments is of several millions per year (at $m_\chi\sim 500$~MeV)
to thousands (at $m_\chi\sim 5$~GeV).  Models with larger masses are
hard to explore.  However, these larger masses correspond to quite
heavy SM Higgs boson, $m_h\sim 350$--$700$~GeV. For the described
model being consistent up to the inflationary scales the Higgs mass
should be $m_h<190$~GeV, and the inflaton mass is below
$m_\chi=1.8$~GeV (see eq.~\eqref{light-inflaton-upper-limit}).  Hence
we conclude that in the model considered here the inflaton sector can
be fully explored in the particle physics experiments.

In this respect it is worth mentioning, that hypothetical light
bosons, (very) weakly coupled to the SM fields, are present in various
extensions of the SM (for particular examples see
\cite{Mahanta:2000zp,Gorbunov:2000th,Gorbunov:2000cz,Pospelov:2008zw}).
Phenomenology of these particles attracted some attention, and in
Ref.~\cite{Amsler:2008zzb} the list of relevant experiments is given
in the section about searches for axions and other very light bosons.
These experiments are relevant for the described model (see the bounds
obtained in this paper).  At the same time the theoretical development
for our case of a light flavour blind scalar dates back to the time,
when very light SM Higgs boson was still allowed experimentally.
Quite a few improvements were done since that time, and we mostly used
in the current work the results obtained in
Refs.~\cite{Ellis:1975ap,Shifman:1979eb,Gasser:1984gg,Leutwyler:1989xj,Donoghue:1990xh}.

The rest of the paper is organised as follows.  In
section~\ref{sec:Model} we describe the model and outline the viable
region in the parameter space, section~\ref{sec:decays} is devoted to
inflaton decays, while meson decays to inflaton are considered in
section~\ref{sec:mesons}. There we obtain lower limits on the inflaton
mass from existing experimental results.  In
section~\ref{sec:production} we study the inflaton production in
$pp$-collisions and give production rates for operating high-intensity
and high-energy beams at JPARC, Fermilab and CERN. Limits on the
inflaton mass from the results of the CHARM experiment are obtained in
section~\ref{sec:long-base-line}. There we also give predictions for
the number of inflatons produced per one year of operation of T2K,
NuMi, CNGS and NuTeV beams in a beam-target setup.
Section~\ref{sec:conclusions} contains conclusions. In
Appendix~\ref{sec:nuMSM} we discuss consequences of obtained results
for a implementation of our model within the
$\nu$MSM~\cite{Shaposhnikov:2006xi}, the extension of the SM with
three sterile right-handed neutrinos, so that the inflaton vacuum
expectation value provides sterile neutrino masses. This model is
capable of explaining neutrino oscillations, dark matter and baryon
asymmetry of the Universe and, equipped with inflaton sector, provides
an example of a full realistic model of particle physics.  Thus the
inflation can be directly tested also in fully realistic extensions of
the SM.

%%%%%%%%%%%%%%%%%%%%%%%%%%%%%%%%%%%%%%%%%%%%%%%%%%%%%%%%%%%%%%%%%%%%%%%%
\section{The model}
\label{sec:Model}

We consider the extension of the SM model with an inflaton field
introduced in \cite{Shaposhnikov:2006xi}.  The Lagrangian of this
extended model reads
\begin{align}
  \L_{X \mathrm{SM}} =
    &\ \L_\mathrm{SM}+\L_{X N}
  \;,\nonumber\\
  \L_{X N} =
    &\ \half \d_\mu X\d^\mu X
  \label{4*} 
     +\half m_X^2 X^2-\frac{\beta}{4} X^4
      - \lambda \l H^\dagger H - \frac{\alpha}{\lambda} X^2 \r^2
  \;,
\end{align}
where $\L_\mathrm{SM}$ is the SM Lagrangian without the Higgs
potential, while the latter gets modified in accordance with
\eqref{4*}, $X$ is a new neutral scalar field and $H$ is the Higgs
doublet.  Note, that it is supposed that the only scale violating term
at tree level is the mass term with the negative squared mass $-m_X^2$
for the extra scalar $X$.  This particular choice is sufficient to
demonstrate the main statement: possibility to test directly the
inflationary models in particle physics experiments.  At the same
time, the algebra below is simpler as compared to those in the case of
general inflaton-Higgs scalar potential, since the number of
parameters is smaller.

The SM-like vacuum of the scalar potential in eq.~\eqref{4*} is 
\begin{gather}
\label{LL}
  \la H \ra = \frac{v}{\sqrt{2}}
  \;,\quad
  \la X \ra = \sqrt{\frac{\lambda}{2\alpha}}\,v
            = \frac{m_\chi}{\sqrt{2\beta}}
  \;,\nonumber\\
  \text{with }v=\sqrt{\frac{2\alpha}{\beta\lambda}}\,m_X =246\GeV
  \;.\label{5*}
\end{gather}
The small excitations about this vacuum have the masses
\begin{equation}\label{5**} 
  m_h=\sqrt{2\,\lambda}\,v
  \;,\quad
  m_\chi=m_h\,\sqrt{\frac{\beta}{2\alpha}}
  \;,
\end{equation}
and are rotated with respect to the gauge basis $(\sqrt{2}H-v,X)$
by the angle
\begin{equation}\label{mixing-angle} 
  \theta=\sqrt{\frac{2\alpha}{\lambda}}=\frac{\sqrt{2\beta}\,v}{m_\chi}
\end{equation}
to the leading order in $\theta$ and $m_\chi/m_h$.

The inflation in the model is supposed to be driven by a flat
potential along the direction
\begin{equation}\label{infl-dir}
  H^\dagger H\cong\frac{\alpha}{\lambda}X^2
  \;.
\end{equation}
The quartic coupling dominates the potential during inflationary and
reheating stages.  Thus, after inflation the Universe expands as at
radiation dominated stage, so the number of required e-foldings is
about $N_e\simeq 62$. The normalisation of the matter power spectrum
\cite{Komatsu:2008hk} generated during inflation fixes\footnote{The
  estimate \eqref{quartic-from-CMB} differs from the other values,
  presented in \cite{Shaposhnikov:2006xi} and in
  \cite{Anisimov:2008qs} due to different values of $N_e$ used there.}
the quartic coupling constant as $\beta\approx \beta_0$, where
\begin{equation}\label{quartic-from-CMB}
  \beta_0=1.5\times 10^{-13}
  \;.
\end{equation}
In this respect it is worth noting, first, that inflation in this
model happens at $\chi\gtrsim M_{Pl}$, where gravity-induced
corrections are expected to be large. We will suppose below, that same
(yet unknown) mechanism, which guarantees the flatness of scalar
potential during inflation, operates at $\chi\lesssim M_{Pl}$ as well.
Thus, the Plank scale physics allows for considering the same scalar
potential \eqref{4*} at high energies with account of perturbative
quantum corrections only due to the gauge and Yukawa couplings. These
latter are discussed in due course.  Second, for the quartic inflation
there is a tension \cite{Komatsu:2008hk} between predicted
tensor-to-scalar amplitudes ratio and fits to cosmological
data. Though the limits \cite{Komatsu:2008hk} are not dramatic yet,
weak non-minimal coupling to Ricci scalar, $\xi X^2R/2$, with $\xi\sim
10^{-3}$ makes the model fully consistent
\cite{Tsujikawa:2004my,Bezrukov:2008dt,Komatsu:2008hk} with all
current cosmological observations.  Switching on $\xi$ results in a
larger value of the quartic coupling constant,
$\beta_0\leq\beta\lesssim2\beta_0$ for $0\leq\xi\lesssim 10^{-3}$.
For our study of inflaton phenomenology at low energies non-minimal
coupling to gravity is irrelevant, but in further estimates we account
for the uncertainty in the value of $\beta$,
\begin{equation}\label{quartic-from-CMB-with-R}
  \beta=\l 1\text{--}2\r \cdot \beta_0
  \;,
\end{equation}
associated with its possible impact.     

Thus, among four parameters in the Lagrangian \eqref{4*}, one,
$\beta$, is fixed by the amplitudes of primordial perturbations and
another combination \eqref{5*} is fixed at the electroweak vacuum by
the value of the Fermi constant.  Two remaining free parameters
determine the SM Higgs boson mass and the inflaton mass.  Further
constraints on them are discussed below.

The baryon asymmetry of the Universe is unexplained within the
framework of the Standard model of particle physics (SM). However, the
baryon number is violated at microscopic level in primordial plasma,
if sphaleron processes are rapid enough \cite{Kuzmin:1985mm}.  This
phenomena is often exploited by mechanisms generating baryon asymmetry
within relevant extensions of the SM. This places a lower bound on the
reheating temperature of the Universe at the level somewhat above the
electroweak scale, and for definiteness we choose $T_r\gtrsim 150$~GeV
(for a review see \cite{Rubakov:1996vz}).  In the model \eqref{4*},
where both quartic coupling and Higgs-to-inflaton mixing are very
weak, but Higgs boson self-coupling is quite strong ($\lambda\gtrsim
0.1$ for $m_h\gtrsim 114$~GeV), the energy transfer from the inflaton
to the SM particles is extremely inefficient
\cite{Anisimov:2008qs}. The stronger is the mixing, the more efficient
is the energy transfer and the higher is the reheat temperature in the
early Universe. Strong mixing (larger $\alpha$) implies lighter
inflaton, see eq.~\eqref{5**}, and lower bound $T_r\gtrsim 150$~GeV
yields \emph{upper bounds} on the inflaton mass
\cite{Anisimov:2008qs}:
\begin{equation}\label{light-inflaton-upper-limit}
  m_\chi\lesssim 1.5\cdot \l \frac{m_h}{150\GeV }\r \cdot 
    \l \frac{\beta}{1.5\times 10^{-13}}\r^{1/2}\,\GeV
  \text{ for } m_\chi < 2 m_h
  \;,
\end{equation}           
and 
\begin{equation}\label{heavy-inflaton-upper-limit}
  m_\chi\lesssim 460 \cdot \l \frac{m_h}{150\GeV }\r^{4/3} \cdot 
    \l \frac{\beta}{1.5\times 10^{-13}}\r^{1/3}\,\GeV
  \text{ for } m_\chi > 2 m_h
  \;. 
\end{equation} 
In the latter case the \emph{lower bound} on the inflaton mass is 
\begin{equation}\label{heavy-inflaton-lower-limit}
  m_\chi > 2 m_h \gtrsim 228\GeV
  \;. 
\end{equation} 
In the former case the \emph{lower bound} on the inflaton mass follows
from the upper limit on the Higgs-inflaton mixing, appearing from the
requirement that quantum corrections, originated from this mixing,
should not dominate over bare coupling constant $\beta$.  With the
action (\ref{4*}) fixed at the electroweak scale the corrections to
the inflationary potential $\beta X^4/4$ can be explicitly calculated
and have the form
\begin{equation}
  \delta V = \frac{m^4(X)}{64\pi^2}\log\frac{m^2(X)}{\mu^2}
  \;,
\end{equation}
where $m(X)$ is the mass of the contributing particle in the inflaton
background field $X$ (taking into account the flat direction
(\ref{infl-dir}) to obtain the Higgs field background, and $\mu$ is
the electroweak scale.\footnote{Strictly speaking this is the
  contribution of a scalar particle, while the numeric coefficient
  changes with the number of spin degrees of freedom, and also changes
  the sign for fermionic particles.}  Then, requiring that in the
inflationary region $X\sim M_P$ the corrections to the quartic
coupling $\beta$ are, somewhat arbitrary, smaller than 10\%, we get
from the contribution of the Higgs boson
\begin{equation}\label{flatness-limit-on-mixing}
  \alpha\lesssim 10^{-7}
  \;,
\end{equation}
which precludes large quantum corrections to inflaton quartic coupling
driving inflation. Limit \eqref{flatness-limit-on-mixing} can be
converted using \eqref{5**} into the lower bound on the inflaton mass
%FB: No alpha should be in the formula!
\begin{equation}\label{light-inflaton-lower-limit}
  m_\chi > 120 \l \frac{m_h}{150\GeV } \r 
  \l \frac{\beta}{1.5\times 10^{-13}}\r^{\frac{1}{2}}
  \l \frac{10^{-7}}{\alpha}\r^{\frac{1}{2}} \MeV
  \;.
\end{equation}
Below this limit one should take into account quantum corrections to
the inflationary potential, which may change the value of the inflaton
coupling constant at the electroweak scale (\ref{quartic-from-CMB}),
or even spoil the inflationary picture.

The proper renormalisation group enhancement of the analysis should be
done, once any experimental evidence for the light inflaton is found.
However, in the larger part of the parameter space no significant
changes to the described bounds are expected.  Limits, similar to
\eqref{flatness-limit-on-mixing}, follow from the requirement of
smallness of the SM gauge and Yukawa coupling corrections.  As far as
all the SM particle masses during inflation are proportional to
$\sqrt{\alpha/\lambda}X$ (see \eqref{infl-dir}), all these bounds
differ from (\ref{flatness-limit-on-mixing}) only by ratios of the
coupling constants of the form $y_t/\sqrt{\lambda}$,
$g/\sqrt{\lambda}$, etc.  This changes the lower bound
(\ref{light-inflaton-lower-limit}).  Note, that the exact value of
this bound is not crucial due to the stronger experimental
constraints, obtained in section \ref{sec:long-base-line}.  However,
in some regions of parameter space these corrections may lead to much
stronger effects---for example, for special ranges of the Higgs boson
mass the Higgs self-coupling becomes small at inflationary scales, so
$\sqrt{\alpha/\lambda}X$ turns to be large, hence so do the SM
perturbative corrections to the inflaton potential.

The weakness of Higgs-to-inflaton mixing
\eqref{flatness-limit-on-mixing} is responsible for very tiny inflaton
interaction with SM particles. This makes searches for heavy inflaton,
in the range given by \eqref{heavy-inflaton-upper-limit},
\eqref{heavy-inflaton-lower-limit}, hopeless in foreseeable
future. Indeed, inflaton direct production requires to collect and
study enormously large statistics in high energy collisions.  In
contrast, the opposite case of light inflaton in the range between
\eqref{light-inflaton-upper-limit} and
\eqref{light-inflaton-lower-limit} is quite promising, since inflaton
production does not require very high energy at a collision point.
Then inflaton can be produced in beam-target experiments, where large
statistics is achievable. In this paper we consider low energy
phenomenology of this light inflaton.

In next Sections we will estimate the decay and production rates of
the light inflaton.  Since inflaton couplings do not depend on the SM
Higgs boson mass, its value determines only viable inflaton mass range
\eqref{light-inflaton-upper-limit},
\eqref{light-inflaton-lower-limit}. Considering for the Higgs boson
mass the range $114\GeV <m_h<700\GeV $ as still possible, in what
follows we study the inflaton low energy phenomenology for its mass in
the interval
\begin{equation}
  \label{Y}
  30\MeV\lesssim m_\chi \lesssim 10\GeV
  \;. 
\end{equation}
Actually, in the model under consideration the upper limit
on the SM Higgs boson mass is lower than 700~GeV, as we use the same
scalar sector \eqref{4*} to describe the inflation at high energies.
Indeed, if one considers the inflationary model \eqref{4*} as it is,
it should be valid (does not become strongly coupled) up to the energy
scale $\sqrt{\alpha/\lambda}X\sim 10^{15}$~GeV.  As far as the
inflaton is very weakly coupled with the Higgs field, this requirement
is the same, as for the Standard Model, leading to the bound
$m_h\lesssim190$~GeV (see, e.g.\ \cite{Bezrukov:2009db}), or
\begin{equation}\label{XXX}
  m_\chi\lesssim1.8~{\rm GeV}
  \;.
\end{equation}
Also, the fit to the electroweak data points at mass interval
$m_h<285$~GeV \cite{Amsler:2008zzb}.  Nevertheless, in extensions of
this model the upper limit on the Higgs boson mass may be higher, so
we will discuss the whole interval \eqref{Y} to make our study
applicable in a more general case.

%%%%%%%%%%%%%%%%%%%%%%%%%%%%%%%%%%%%%%%%%%%%%%%%%%%%%%%%%%%%%%%%%%%%%%%%
\section{Inflaton decay palette}
\label{sec:decays}

Light inflaton decays due to the mixing with the SM Higgs boson
\eqref{mixing-angle}.  Thus, its branching ratios coincide (taking
into account the small mixing angle (\ref{mixing-angle})) with those
of the light SM Higgs boson, studied in \cite{Ellis:1975ap}.  In what
follows we actually update the results of \cite{Ellis:1975ap} in view
of further relevant developments and findings.

Inflaton of the mass below 900~MeV decays mostly into 
\[
  \gamma\gamma,\;e^+e^-,\; \mu^+\mu^-,\;\pi^+\pi^-,\;\pi^0\pi^0
  \;.
\]
Mixing \eqref{mixing-angle} gives rise to the Yukawa-type inflaton coupling
to the SM fermions $f$,
\begin{equation}
  \label{inflaton-to-fermions-coupling}
  \L_{\chi\bar f f}=\theta\,\frac{m_f}{v}\, \chi \bar f f = \sqrt{2\beta}\,
  \frac{m_f}{m_\chi} \chi \bar f f
  \;,
\end{equation}
where $m_f$ is the fermion mass.  Effective inflaton-pion interaction
follows from the Higgs boson coupling to the trace of the
energy-momentum tensor, (cf.~\cite{Leutwyler:1989xj,Gasser:1984gg})
\begin{align}
  \label{inflaton-to-pions-coupling}
  \L_{\chi\pi\pi}=
    &~ 2\kappa\sqrt{2\beta}\cdot\frac{\chi}{m_\chi}\cdot 
      \l 
        \half \d_\mu\pi^0\d^\mu\pi^0+\d_\mu\pi^+\d^\mu\pi^-
      \r
  \\\nonumber
    &-(3\kappa+1)\sqrt{2\beta}\cdot\frac{\chi}{m_\chi}\cdot m_\pi^2 
     \cdot \l 
       \half \pi^0\pi^0+\pi^+\pi^-
     \r
  \;,
\end{align}
with $m_\pi$ being the pion mass and
\begin{equation}
  \label{eq:1}
  \kappa = \frac{2N_h}{3b} = \frac{2}{9}
  \;,
\end{equation}
where $N_h=3$ is the number of heavy flavours, $b=9$ is the first
coefficient in the QCD beta function without heavy quarks.  Finally,
Higgs-inflaton mixing results in the inflaton coupling to $W$-boson,
which contributes to the triangle one-loop diagram responsible for the
inflaton decay to a pair of photons.  Similar contributions come from
fermion loops, so that inflaton decay to photons can be described by
the effective Lagrangian (cf.~\cite{Shifman:1979eb})
\begin{equation}
  \label{inflaton-to-photons-coupling}
  \L_{\chi\gamma\gamma}\approx \frac{F\alpha}{4\pi}
  \,\frac{\sqrt{2\beta}}{m_\chi}\, \chi\,
  F_{\mu\nu}F^{\mu\nu}
  \;,    
\end{equation}
where $\alpha$ is the fine structure constant, $F$ is a sum of loop
contributions from $W$-boson and fermions $f$ with electric charge
$eq_f$, $F=F_W+\sum_{f,\mathrm{colors}} q_f^2 F_f$
\begin{align}
  \label{ccc}
  F_W &=2+3y\left[1+ \l 2-y\r x^2\right] 
  \;,\\
  F_f &= -2 y \left[1+ \l 1-y\r x^2\right]
  \;,\nonumber
\end{align}
and $y=4 m^2/m_\chi^2$, with $m$ being the mass of the contributing
particle; here also
\begin{align*}
  x=\mathrm{Arctan}\frac{1}{\sqrt{y-1}}
  \;,& \text{ for } y>1
  \;,\\
  x=\half \l \pi + i \, \log \frac{1+\sqrt{1-y}}{1-\sqrt{1-y}}\r
  \;,& \text{ for } y<1
  \;. 
\end{align*}
In fact, for the interesting range of the inflaton mass the fermion
contributions almost cancel the $W$-boson contribution (the latter
dominates over a contribution of each single fermion).

The inflaton-to-SM fields couplings presented above yield the inflaton
decays to leptons with the rates:
\begin{equation}\label{eq:gammall}
  \Gamma_{\chi \to \bar l l}
%    = \xi^2\frac{m_\chi}{8\pi} \frac{m_l^2}{v^2}
%      \l 1-\frac{4m_f^2}{m_I^2} \r^{3/2}
%
  = \frac{\beta m_l}{4\pi}\frac{m_l}{m_\chi}
  \l 1-\frac{4m_f^2}{m_\chi^2} \r^{3/2}
  \;,
\end{equation}
the inflaton decays to pions with the rates:  
\begin{equation}\label{decay-to-pions}
  \Gamma_{\chi \to \pi^+ \pi^-}=2 \Gamma_{\chi \to \pi^0 \pi^0}= 
%  \frac{\xi^2m_\chi^3}{16\pi v^2}\cdot \l
%  \frac{2}{9}+\frac{11}{9}\frac{m_\pi^2}{m_\chi^2}\r^2 
%  \sqrt{1-\frac{4m_\pi^2}{m_\chi^2}}
%  =
  \frac{\beta m_\chi}{8\pi}\cdot \l
  \frac{2}{9}+\frac{11}{9}\frac{m_\pi^2}{m_\chi^2}\r^2 
  \sqrt{1-\frac{4m_\pi^2}{m_\chi^2}}
  \;,
\end{equation}
and the inflaton decay to photons with the rate:
\begin{equation}
  \label{decay-to-photons}
  \Gamma_{\chi\to \gamma\gamma}
%    = \xi^2\frac{m_\chi}{8\pi} \frac{m_l^2}{v^2}
%      \l 1-\frac{4m_f^2}{m_I^2} \r^{3/2}
    = \left| F \right|^2\, \l\frac{ \alpha}{4\pi}\r^2
  \frac{\beta\, m_\chi}{8\pi}
  \,.
\end{equation}
Note, the tree-level estimate \eqref{decay-to-pions} is not correct
far from the pion threshold, where strong final state interactions
become important \cite{Donoghue:1990xh}. Thus, in our numerical
estimates we follow Ref.~\cite{Donoghue:1990xh} to improve
\eqref{decay-to-pions}.

If inflaton is heavier than 900~MeV, its hadronic decay modes become
more complicated.  First, each quark flavour contributes to the
inflaton decay if the inflaton mass exceeds the double mass of the
lightest hadron of the corresponding flavour.\footnote{We do not
  discuss decays below thresholds with one hadron being off-shell.}
Strange quark starts to contribute for the inflaton mass $m_\chi>2
m_K$, where $m_K$ is the $K$-meson mass. Charm quarks contribute, if
$m_\chi> 2 m_D$, where $m_D$ is the $D$-meson mass. Lightest flavour
states are mesons and close to the thresholds the inflaton decays are
described by effective interactions similar to
\eqref{inflaton-to-pions-coupling}. Farther from the thresholds final
state interactions become important and we follow
Ref.~\cite{Donoghue:1990xh} to estimate the decay branching rate to
pions and kaons. This approach is valid until other flavour resonances
enter the game or while multimeson final states are negligible. For
pions this approach becomes unjustified in models with inflaton mass
above approximately 1~GeV\@.  For mesons of heavier flavours the
approach fails not far from the corresponding thresholds, where
produced mesons are not relativistic but many new hadrons enter the
game. Far above a quark threshold inflaton decay can be described as
decay into a pair of quarks due to coupling
\eqref{inflaton-to-fermions-coupling}, which subsequently
hadronise. To estimate the decay rate, QCD-corrections should be
accounted for, which are the same as those for the SM Higgs boson of
mass $m_\chi$.

Second, if the inflaton mass is close to the mass of some narrow
hadronic scalar resonance (e.g., oniums in $c\bar c$ and $b\bar b$
systems), then interference with such a resonance changes
significantly inflaton hadronic decay rates in a way quite similar to
what was expected for the SM Higgs boson if its mass would be close to
the $b\bar b$ threshold \cite{Ellis:1979jy}. We do not consider here
this case, though the calculations are straightforward.

Third, for the inflaton mass above 1~GeV the decay into mesons
actively proceeds via the inflaton-gluon-gluon coupling, which due to
the quark one-loop triangle diagram similar to that contributing to
the inflaton decays to photons. The effective Lagrangian describing
this decay is \cite{Ellis:1979jy}
\begin{equation}\label{inflaton-to-gluons-coupling}
  \L_{\chi gg} \approx \frac{F\alpha_s}{4\sqrt{8}\,\pi}
  \,\frac{\sqrt{2\beta}}{m_\chi}\, \chi\,
  G^a_{\mu\nu}G^{a\,\mu\nu}
  \;,
\end{equation}
where $\alpha_s$ is the strong gauge coupling constant and $F=\sum_f
F_f$, see \eqref{ccc}, with sum only over quarks. Higher order QCD
corrections are also important here, and they coincide with those in
the case of the SM Higgs boson. The produced gluons hadronise later
on.

Obviously, well above the heavy quark threshold and well above QCD
scale the description of the inflaton decays in terms of quarks and
gluons is well-justified, while in the opposite case the effective
description in terms of mesons is applicable.  Between these ranges,
where $m_\chi\simeq 1.5$--$2.5$~GeV, both approximations are not quite
correct and no reliable description can be presented. At the same
time, comparing relevant hadronic contributions described within these
two approaches valid at lower and upper limits of this ``untractable''
interval, we observe deviations not larger than by an order of
magnitude. Thus we conclude, that order-of-magnitude estimates of
lifetime and decay rates of leptonic, photonic and total hadronic
modes for the inflaton mass interval 1.5--2.5~GeV can be obtained by
some interpolation between these ``low-mass'' and ``high-mass''
results.

Hence, for the inflaton decay rates to quarks we obtain the same
formula as \eqref{eq:gammall} multiplied by the number of colours, 3,
and by a factor due to QCD corrections (having in mind uncertainties
in the value of $\beta$ discussed in section~\ref{sec:Model}, for our
estimates we adopt only leading order QCD corrections presented in
\cite{Spira:1997dg}).  For decay rates into gluons we obtain
\[
  \Gamma_{\chi\to gg} =   \left| F \right|^2\, \l\frac{ \alpha_s}{4\pi}\r^2
  \frac{\beta\, m_\chi}{4\pi}
  \;,
\]  
multiplied by the corresponding factor due to leading-order
QCD-corrections \cite{Spira:1997dg}.

%\begin{figure}
\FIGURE{
  \centerline{%
    \begin{tabular}{c}
      \includegraphics{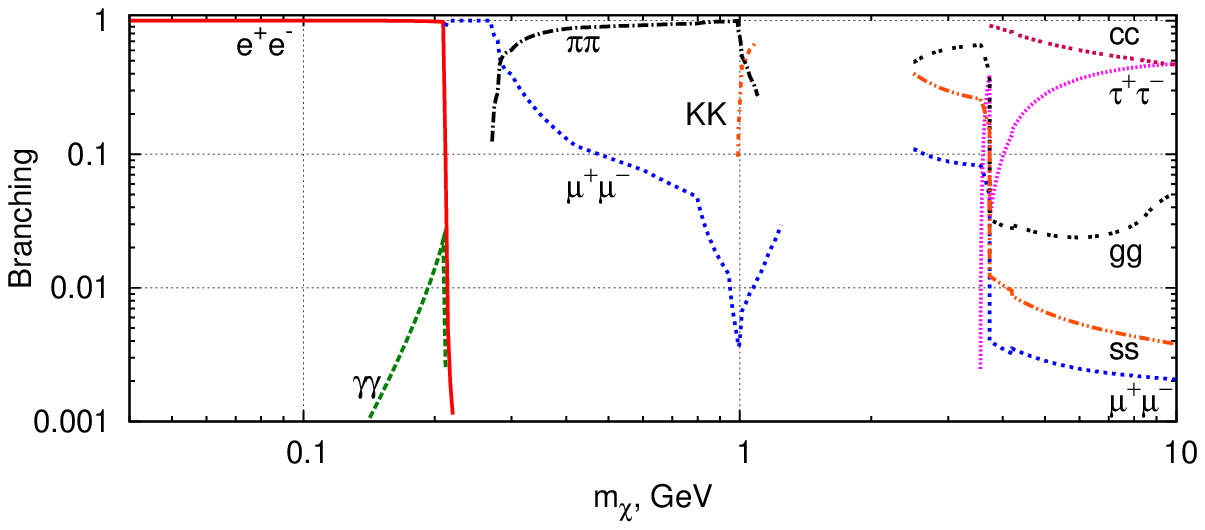} \\
      \includegraphics{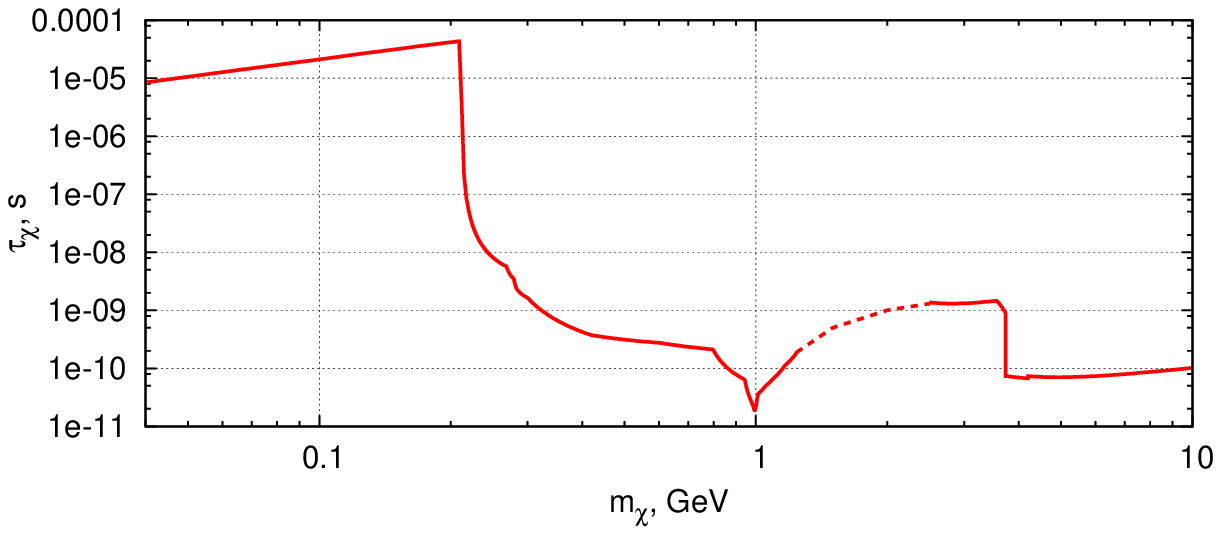}
    \end{tabular}}
  \caption{\emph{Upper panel:}
    Inflaton branching ratios to various two-body final states 
    as functions of inflaton mass
    $m_\chi$, for 1.5-2.5~GeV range see discussion in the main text; 
     \emph{Lower panel:}  
     inflaton lifetime $\tau_\chi$ as a function of the inflaton
    mass $m_\chi$.  Here we set $\beta=\beta_0$, while for other
    values the life-time is
    inversely proportional to $\beta$.  With $\beta$ within
    the favourable interval \protect\eqref{quartic-from-CMB-with-R} the
    lifetime can be two times smaller.
  }
  \label{infdecays}
}
%\end{figure}

The inflaton branching ratios and lifetime are presented in
Fig.~\ref{infdecays} as functions of the inflaton mass.  Note, that
inflaton partial widths into fermions decrease with increasing
inflaton mass, while the opposite behaviour is observed for decays
into pions due to the contribution of kinetic term from the trace of
the energy-momentum. Diphoton mode varies with the inflaton mass and
for the interesting range of parameters reaches its maximum of about
2.5\% just at muon threshold.  Thus, the sub-GeV inflaton
predominantly decays into electrons, if its mass is below 200~MeV,
otherwise to muons and pions with comparable branching ratios. If the
inflaton mass is below the muon threshold, its life-time is about
$10^{-5}$~s and above this threshold it falls rapidly down to
$10^{-9}$~s.  For the inflaton with mass well above 1~GeV, heaviest
kinematically allowed fermions dominate the inflaton decay.  Decays
into photons and electrons are strongly suppressed for $m_\chi\gtrsim
1$~GeV, with branchings below $10^{-4}$.

%%%%%%%%%%%%%%%%%%%%%%%%%%%%%%%%%%%%%%%%%%%%%%%%%%%%%%%%%%%%%%%%%%%%%%%%
\section{Inflaton from hadron decays}
\label{sec:mesons}

In this section we consider the inflaton production in rare decays of
mesons and obtain corresponding limits on the light inflaton mass.

First, light inflaton can be produced in two-body meson decays. These
are exactly the processes widely discussed in the past, when the SM
Higgs boson was considered a (sub)GeV particle.  Making use of the
results from \cite{Leutwyler:1989xj} one obtains for the amplitudes of
the kaon decays
\begin{align}\label{charged-kaon-decay}
  A\l K^+\to \pi^+\chi\r
    & \simeq
      \theta\frac{M_K^2}{v}\Bigg\{
        \gamma_1\frac{1-\kappa}{2}\left(1-\frac{m_\chi^2-M_\pi^2}{M_K^2}\right)
  \\\nonumber
    & \hphantom{\simeq
      \theta\frac{M_K^2}{v}\Bigg\{}
        -\gamma_2(1-\kappa)
        +\frac{1}{2}
        \frac{3G_F\sqrt{2}}{16\pi^2} 
        \sum_{i=c,t} V_{id}^* m_i^2 V_{is}\; 
      \Bigg\}
  \;, \\
  A\l K_L\to \pi^0\chi\r
    & = - A\l K^+ \to \pi^+ \chi\r
  \;, \\
  A\l K_S\to \chi\pi^0\r
    & \simeq \theta\frac{3G_F\sqrt{2}}{16\pi^2} 
      \frac{iM_K^2}{2 v}\, {\Imag}\sum_{i=c,t} V_{id}^* m_i^2 V_{is}
  \;,
\end{align}
with $\gamma_1\sim3.1\times10^{-7}$ and much smaller $\gamma_2$.  The
largest contribution comes from the third term in
\eqref{charged-kaon-decay}.  At quark level this term is due to the
inflaton emission by a virtual quark in the quark-W-boson loop.  For
the branching ratios we get
\begin{align}
    \Br\l K^+ \to \pi^+ \chi\r
    &= \frac{1}{\Gamma_\mathrm{total}(K^+)}
    \frac{|A(K^+\to\pi^+\chi)|^2}{16\pi M_K}\frac{2|\mathbf{p}_\chi|}{M_K}
  \nonumber\\
  \label{BrKpih}
    & \approx
      1.3\times 10^{-3}\cdot \l \frac{2|\mathbf{p}_\chi|}{M_K} \r\theta^2
  \\
    & \approx 2.3\times 10^{-9}\cdot \l \frac{2|\mathbf{p}_\chi|}{M_K} \r
      \cdot \l \frac{\beta}{\beta_0}\r \cdot \l
      \frac{100\MeV}{m_\chi}\r^2
  \;,\nonumber\\
  \label{BrKLpih}
    \Br\l K_L \to \pi^0 \chi\r
    & \approx 5.5\times 10^{-3}\cdot \l \frac{2|\mathbf{p}_\chi|}{M_K} \r\theta^2
  \nonumber\\
    & \approx 1.0\times 10^{-8}\cdot \l \frac{2|\mathbf{p}_\chi|}{M_K} \r
    \cdot \l \frac{\beta}{\beta_0}\r \cdot \l
    \frac{100\MeV}{m_\chi}\r^2
  \;,
\end{align}
where $\mathbf{p}_\chi$ is the inflaton 3-momentum.  The branching
ratio of $K_S$ is much smaller.  For the inflaton mass in the
kinematically allowed range the squared mixing angle $\theta^2$ is of
the order $10^{-5}$--$10^{-7}$ \eqref{mixing-angle}\@.  For a model with
$\beta =\beta_0$ branching ratios of the kaon decays are presented in
Fig.~\ref{fig:Kaons} together with the relevant existing limits from
the searches of the processes $K\to \pi + \text{nothing}$
\cite{Adler:2004hp,Artamonov:2009sz}.  It follows from
Fig.~\ref{fig:Kaons} that the models with light inflatons,
\[
  m_\chi\lesssim 120\MeV
  \;,
\]
are \emph{excluded} by negative results of these searches and models
with $170\MeV\lesssim m_\chi \lesssim 205$~MeV are disfavoured.
Further increase in sensitivity of these searches by one order of
magnitude would allow to explore the light inflaton in the mass region
150--250~MeV. Thus, kaon decays are the most promising processes to
search for the light inflaton.

%\begin{figure}
\FIGURE{
  \centerline{\includegraphics{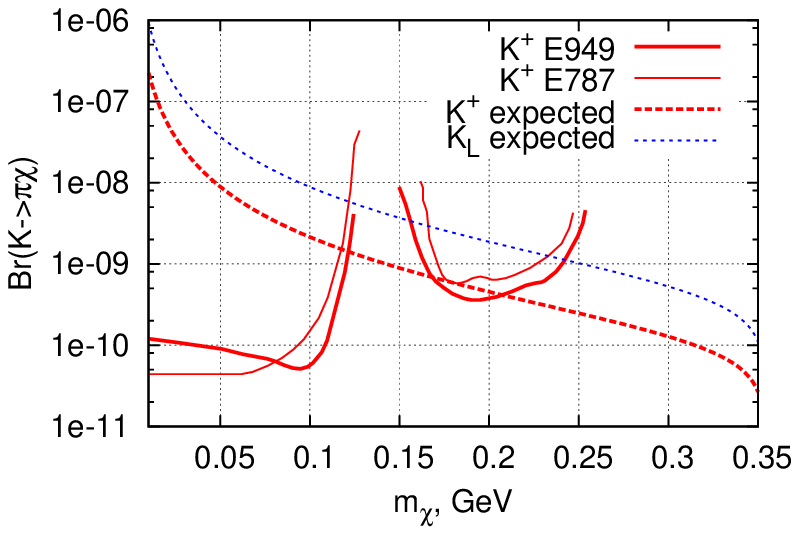}}
    \caption{Expected branching ratios $\Br\l K^+ \to \pi^+ \chi\r$,
      $ \Br\l K_L \to \pi^0 \chi\r $ and experimental bounds on $
      \Br\l K^+ \to \pi^+ \chi\r$ from \cite{Artamonov:2009sz} and
      \cite{Adler:2004hp}.}
    \label{fig:Kaons}
}
%\end{figure}

In case of a larger inflaton mass heavy mesons have to be considered.
The most promising here is the $\eta$-meson with the branching ratio
of the order (cf.\ \cite{Leutwyler:1989xj,Kozlov:1995yd})
\begin{equation}\label{eta-decay}
  \Br\l \eta \to \chi\pi^0\r \sim
  \frac{|\mathbf{p}_\chi|}{M_\eta}\cdot10^{-6}\cdot\theta^2 \approx 
  1.8\times 10^{-12}\cdot \l \frac{2|\mathbf{p}_\chi|}{M_\eta} \r
  \cdot \l \frac{\beta}{\beta_0}\r \cdot \l \frac{100\MeV}{m_\chi}\r^2
  \;,
\end{equation}
while the branching ratios of vector mesons are about two orders of
magnitude smaller (cf.\ \cite{Paver:1989ct}).

Two-body decays of charmed mesons similar to $K\to \pi \chi$ are
strongly suppressed as compared to that decay because of the smallness
of the up-quark mass and the corresponding CKM matrix elements in the
amplitude (cf.\ the third term in \eqref{charged-kaon-decay} which
dominates).  Three-body semileptonic decays have larger rates, however
they are still quite small \cite{Dawson:1989kr,Cheng:1989ib},
\begin{align*}
  \Br\l D \to e\nu\chi\r
    & =
      \frac{\sqrt{2}G_F m_D^4}{96\pi^2 m_\mu^2 (1-m_\mu^2/m_D^2)^2}
      \cdot \frac{7}{9} \theta^2
      \cdot \Br(D\to\mu\nu) f\left(\frac{m_\chi^2}{m_D^2}\right)
  \\
    & \approx
      5.7 \times 10^{-9} \theta^2 f\left(\frac{m_\chi^2}{m_D^2}\right)
  \;,\\
  f(x) &= (1-8x+x^2)(1-x^2)-12x^2\log(x)
  \;.
\end{align*}
Similar formula for the $D\to\mu\nu\chi$ decay rate is a bit more
complicated because of the larger muon mass \cite{Dawson:1989kr}, but
the rate is suppressed equally strong.  On the contrary, decays of the
beauty mesons are enhanced as compared to $K\to \pi \chi$.  From
Ref.~\cite{Chivukula:1988gp,Grinstein:1988yu} and
\eqref{mixing-angle},\eqref{quartic-from-CMB} we obtain for the light
inflaton:
\begin{align}
  \label{BrhX}
  \Br\l B\to \chi X_s\r
    & \simeq 0.3 \frac{\left|
      V_{ts}V_{tb}^*\right|^2}{\left| V_{cb}\right|^2}  \l \frac{m_t}{M_W}
      \r^4 \l 1-\frac{m_\chi^2}{m_b^2}\r^2 \theta^2
  \\\nonumber
    & \simeq 10^{-6} \cdot\l
      1-\frac{m_\chi^2}{m_b^2}\r^2 \l \frac{\beta}{\beta_0}\r \l
      \frac{300\MeV}{m_\chi}\r^2
  \;,
\end{align}
where $X_s$ stands for strange meson channel mostly saturated by a sum
of pseudoscalar and vector kaons.  The inflaton with the mass below
the muon threshold escapes the detector (see Fig.~\ref{infdecays})
giving the signatures
\[
  B\to K+\text{nothing}
  \;,\qquad
  B\to K^*+\text{nothing}
  \;.
\] 
Heavier inflaton can decay within the detector, with the most clear
mode being the muon one at the level of $0.01$--$1$, depending on the
mass, see Fig.~\ref{infdecays}\@.  Note, that collected world
statistics at b-factories allowed to measure branching fractions of
relevant decays $B\to K^{(*)}l^+ l^-$ with accuracy of about
$10^{-7}$~\cite{BELLE:2009zv}, which is comparable to the expected
signal \eqref{BrhX}.  Thus, an appropriate reanalysis of these data
might give a chance to probe the inflaton of mass $m_\chi\sim
300$~MeV.

Inflaton can also be produced in other meson decays and heavy baryon
decays. We do not discuss these channels considering them as
subdominant for the inflaton production and less promising in direct
searches for the light inflaton.

%%%%%%%%%%%%%%%%%%%%%%%%%%%%%%%%%%%%%%%%%%%%%%%%%%%%%%%%%%%%%%%%%%%%%%%%
\section{Inflaton production in particle collisions}
\label{sec:production}

In this Section we discuss in detail the inflaton production in
particle collisions.  If the collision energy is large enough, the
most efficient mechanism of inflaton production is kinematically
allowed decays of heavy mesons, produced in the collision.  Then the
production cross section $\sigma$ can be estimated as the product of
the meson production cross section and the branching of the meson
decay into the inflaton,
\begin{equation}
  \label{MesonProduction}
  \frac{\sigma}{\sigma_{pp,\mathrm{total}}} =
  M_{pp} \big(
    \chi_s (0.5\Br(K^+\to\pi^+\chi)+0.25\Br(K_L\to\pi^0\chi))
    +\chi_c \Br(B\to\chi X_s)
  \big)
  \;,
\end{equation}
where total hadron multiplicity $M_{pp}$ and relative parts going into
different flavours $\chi_{s,c}$ are given in Table~\ref{beams-details}
for several existing beams, and $\sigma_{pp,\mathrm{total}}\simeq
40$~mbarn is the total proton cross section \cite{Amsler:2008zzb}.
Here only strange and beauty mesons were taken into account, as they
give the main contribution.  In Fig.~\ref{fig:sigmaprod} we present
the estimate of this indirect inflaton production in a beam-target
experiment for several available proton beams.

%\begin{table}[!htb]
\TABLE{
  \centerline{%
  \begin{tabular}{|c||c|c|c|c|c|c|}
    \hline 
    Experiment
      & $E$, GeV
            & $N_{POT}$, $10^{19}$
                  & $M_{pp}$~\cite{Amsler:2008zzb}
                       & $\chi_s$~\cite{Andersson:1983ia,Bowler:1981sb}
                               & $\chi_c$~\cite{Lourenco:2006vw}
                                                     &
                                                $\chi_b$~\cite{Lourenco:2006vw} 
    \\\hline
    CNGS~\cite{CNGS}
      & 400 & 4.5 & 13 & $1/7$ & $0.45\cdot 10^{-3}$ & $3\cdot 10^{-8}$
    \\
    NuMi~\cite{NuMi}
      & 120 & 5   & 11 & $1/7$ & $1\cdot 10^{-4}$    & $10^{-10}$
    \\
    T2K~\cite{T2K}
      & 50  & 100 & 7  & $1/7$ & $1\cdot 10^{-5}$    & $10^{-12}$
    \\
    NuTeV~\cite{NuTeV}
      & 800 & 1   & 15 & $1/7$ & $1\cdot 10^{-3}$    & $2\cdot 10^{-7}$
    \\\hline
  \end{tabular}}
  \caption{Adopted values of relevant for inflaton production parameters of
    several experiments.
    \label{beams-details}}
}
%\end{table}

%\begin{figure}[!htb]
\FIGURE{
  \centerline{\includegraphics{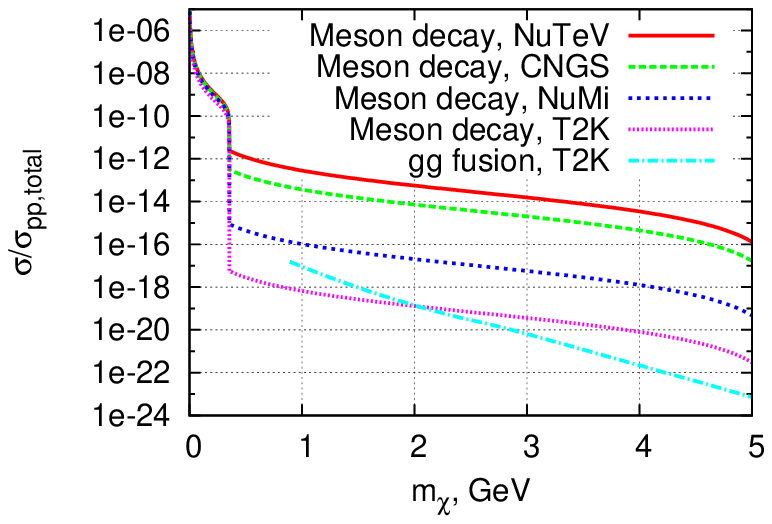}}
  \caption{The inflaton production cross section in $pp$ collisions
    normalised to the total $pp$ cross section $\sigma_{pp\
    total}\simeq 
    40$~mbarn.  For all graphs $\beta=1.5\times10^{-13}$.  The lines are for
    several existing beam energies (see Table~\protect\ref{beams-details}).
    For the T2K case the estimate of the direct production by gluon
    fusion is also given, while for higher beam energies it is negligible,
    compared to the meson decay channel of production.  At the kaon
    threshold $m_\chi\sim m_K$ the account of the contribution from the
    $\eta$-meson decay \protect\eqref{eta-decay} might somehow smooth the
    cross section change.
  \label{fig:sigmaprod}}
}
%\end{figure}

In the models with the inflaton mass above the bottom quark threshold,
the dominant source of the inflatons is the direct production in hard
processes similar to the case of the SM Higgs boson.  The main channel
for the inflaton is the same as for the Higgs boson, i.e.\ the
gluon-gluon fusion \cite{Georgi:1977gs}.  The inflaton production is
calculated exactly as the production of the Higgs boson of the same
mass, only it is suppressed by the mixing angle squared $\theta^2$.
For several available proton beams the result of the calculation with
the parton distribution functions from \cite{Alekhin2006} is given in
Fig.~\ref{fig:disproduction}.  Here we present the estimate of the
direct contribution to the inflaton production even for small inflaton
mass down to $m_\chi\gtrsim 2$~GeV, to illustrate the statement that
indirect production dominates below the B-meson threshold, cf.\
Fig.~\ref{fig:sigmaprod}.

%\begin{figure}[!htb]
\FIGURE{
  \centerline{\includegraphics{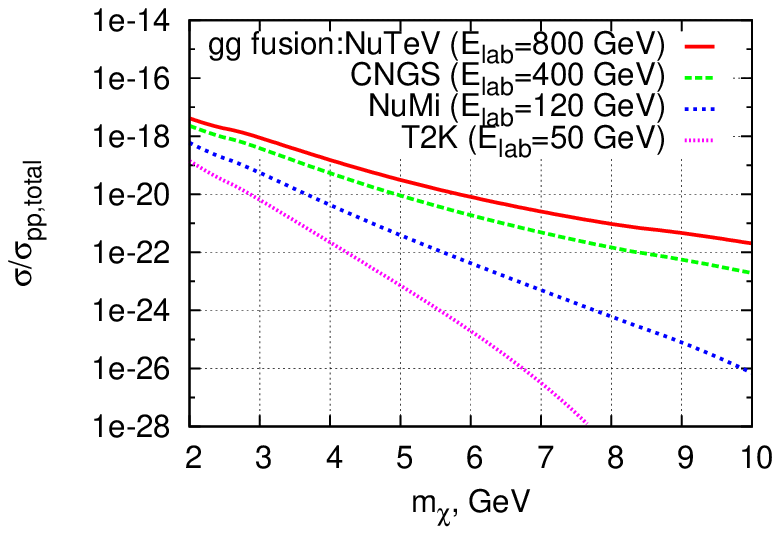}}
  \caption{\label{fig:disproduction} The ratio of the inflaton direct
    production cross section in $pp$ collisions to the total $pp$
    cross section, $\sigma_{pp\ total}\sim 40$~mbarn, for inflationary
    $\beta=\beta_0$.
    %The dots indicate the uncertainties estimated by
    %varying renormalization scale $\mu$ between $m_\chi/2$ and
    %$2m_\chi$.
  }
}
%\end{figure}

%%%%%%%%%%%%%%%%%%%%%%%%%%%%%%%%%%%%%%%%%%%%%%%%%%%%%%%%%%%%%%%%%%%%%%%%
\section{Limits from direct searches and predictions for forthcoming
  experiments}
\label{sec:long-base-line}

The light inflaton will be produced in large amounts in any beam dump,
and then decay further into a photon or lepton pair, depending on its
mass.  Search of a penetrating particle of this type was performed by
the CHARM experiment \cite{Bergsma:1985qz}.  In this article the
search for a generic axion was performed.  However, the only
difference between the axion in \cite{Bergsma:1985qz} and the light
inflaton is in the estimate of the decay rate and production cross
section.  Here we will make rough reanalysis for the case of the light
inflaton.  As far as in the most interesting region of masses the
inflaton lifetime is small and a significant amount of inflatons decay
before they reach the detector, this reanalysis can not be made by
simple rescaling of the result.  So, first we will reproduce in the
simplest possible way the resulting bound from \cite{Bergsma:1985qz},
and then change in the analysis the decay and production rates.  The
resulting picture in \cite{Bergsma:1985qz} can be reproduced by
demanding that the number of decays in the detector is larger than the
background.  The number of decays in the detector is roughly estimated
as
\begin{equation}\label{eq:CHARMaxion}
  N \simeq
  N_0 \sigma_X
  \e^{-\Gamma \frac{l_{dec}}{\gamma}}
  \left(1-\e^{-\Gamma \frac{l_{detector}}{\gamma}}\right)
  \;,
\end{equation}
where $N_0$ is the overall coefficient describing luminosity,
$\sigma_X$ is the production cross section of the axion (eq.~(5) in
\cite{Bergsma:1985qz}), $\Gamma$ is the decay width (sum of eqs.~(3)
and (4) for muons and electrons in \cite{Bergsma:1985qz}),
$l_{dec}=480$~m is the decay length before the detector,
$l_{detector}=35$~m is the detector length, $\gamma=E/m_X$ is the
typical relativistic gamma factor of the axion, with $E\sim 10\GeV$.
Then, Fig.~4 from \cite{Bergsma:1985qz} is approximately reproduced
for $N/(N_0 \sigma_{\pi^0})\simeq 10^{-17}$.

Using the same logic we can obtain the bound for the inflaton.  We
then get instead of (\ref{eq:CHARMaxion})
\begin{multline}
  \label{eq:3}
  N \simeq
  N_0 \sigma_{prod}
  (\Br(\chi\to\gamma\gamma)+\Br(\chi\to ee)+\Br(\chi\to\mu\mu))
  \\\times
  \e^{-\Gamma \frac{l_{dec}}{c\gamma}}
  \left(1-\e^{-\Gamma \frac{l_{detector}}{c\gamma}}\right)
  \;,  
\end{multline}
where for the inflaton production cross section $\sigma_\chi$ we take
\eqref{MesonProduction}, and we adopt the simple estimate of the
$\pi_0$ yield: $\sigma_{\pi^0}/\sigma_{pp,\mathrm{total}}=M_{pp}/3$,
with $M_{pp}=13$ for the CNGS beam (see Table~\ref{beams-details}).
Then, the region forbidden by the experiment \cite{Bergsma:1985qz} is
given in Fig.~\ref{fig:charm}.

%\begin{figure}[!htb]
\FIGURE{
  \centerline{\includegraphics{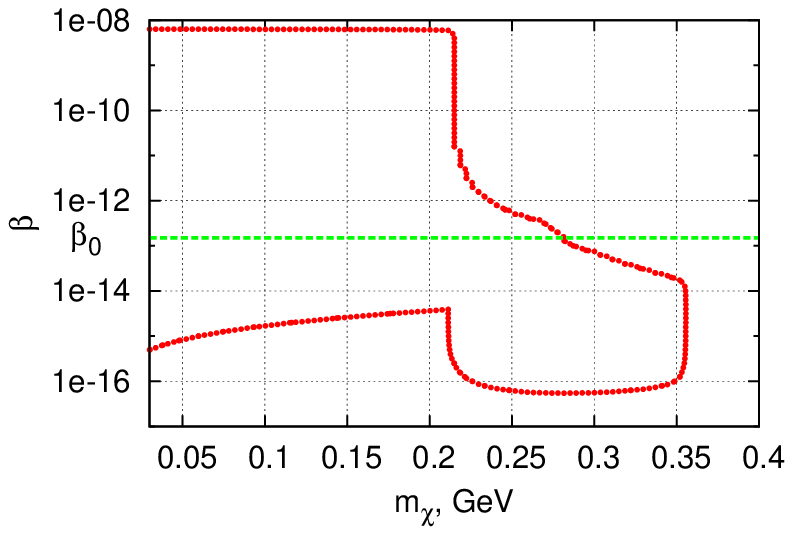}}
  \caption{Limits on the inflaton properties from the CHARM axion searches.
    The region of $\l m_\chi,\beta\r$ surrounded by the curve
    is forbidden. 
    One can see, that for inflationary $\beta=\beta_0$ the 
    inflatons with masses up to approximately 280~MeV are ruled out.
    \label{fig:charm}}
}
%\end{figure}

One can see, that for the inflationary self coupling $\beta=\beta_0$
the CHARM results exclude the inflaton with masses $m_\chi\gtrsim
280$~MeV, while for the upper limit of the interval
\eqref{quartic-from-CMB-with-R} we have
\begin{equation}\label{LLL}
  m_\chi \gtrsim 270\MeV
  \;.
\end{equation}
One should be somewhat careful taking this limit.  The exact
dependence of the bound on $\beta$ in this region is sensitive to the
proper analysis of the CHARM experiment (proper simulation of energy
distribution of the produced inflatons, detector sensitivity to
different decay modes, etc.).  From the plot in Fig.~\ref{fig:charm}
we conclude, however, that the value of $m_\chi>210\MeV$ is a
conservative bound: within the simple approach above we get an order
of magnitude ``safety margin''.  The definite conclusion about higher
values of masses for $\beta\sim\beta_0$ requires careful reanalysis of
the CHARM data.

For the allowed inflaton mass range and using the results obtained in
section~\ref{sec:production} we estimate the number of inflatons
produced during one year of running at designed luminosity for several
experiments by multiplying the cross section ratio by the number of
protons on target, see Fig.~\ref{fig:Nproduced} and
Table~\ref{beams-details}.  This number should be taken with a grain
of salt, as far as we have totally neglected all possible geometrical
factors and possible collimation or deflection of the produced charged
kaons. However one can conclude that higher energy beams are certainly
preferable in the searches for the inflaton, because the dominant
production mode is decays of beauty hadrons.\footnote{This is not so
  for low mass inflatons, which can be produced in kaon decays and one
  can win slightly because of usually higher luminosity of the lower
  energy beam.}  For small mass the number of the inflatons per year
can exceed several millions, while at $m_\chi\sim 5$~GeV it is about a
thousand at best.

%\begin{figure}[!htb]
\FIGURE{
 \centerline{\includegraphics{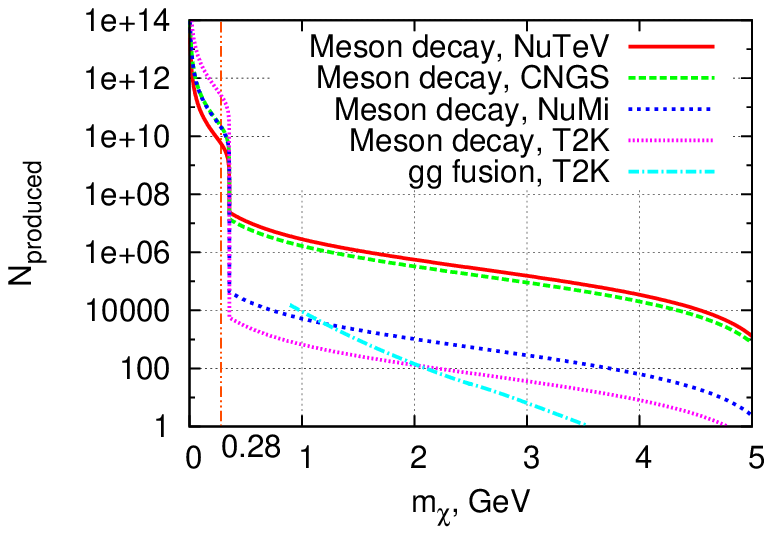}}
  \caption{Number of the inflatons produced with one-year statistics for the
    reference beams.  All geometric factors are neglected. The
    vertical line indicates the lower limit on the inflaton mass
    protect\protect\eqref{LLL}. 
    \label{fig:Nproduced}}
}
%\end{figure}

%%%%%%%%%%%%%%%%%%%%%%%%%%%%%%%%%%%%%%%%%%%%%%%%%%%%%%%%%%%%%%%%%%%%%%%%
\section{Conclusions}
\label{sec:conclusions}

In this paper we presented an example of a simple
inflationary model, which can be fully explored in particle physics
experiments. The inflaton is light and can be hunted for in the
beam-target experiments and in two-body meson decays. The inflaton
life-time is about $10^{-9}$--$10^{-10}$~s, so it decays close to the
production point. The dominant mechanism of inflaton production is the
meson decays. Thus, and it is quite amusing in fact, the first stage
of the Universe evolution described by means of quantum field theory
can be directly tested with modern experimental techniques.

The mass limits on the light inflaton are found to be $270\MeV\lesssim
m_\chi\lesssim 1.8$~GeV.  Note, that careful reanalysis of the results
of the CHARM experiment \cite{Bergsma:1985qz} may change the lower
bound given here.

The paper analysed the quartic inflaton self-interaction with minimal
(or very weakly non-minimal) coupling to gravity, which is still
allowed by the current WMAP data.  In near future results from the
PLANCK experiment will determine the inflationary
parameters---spectral index $n_s$ and tensor-to-scalar ratio
$r$---with much better precision.  This will allow to fix the value of
the non-minimal coupling constant $\xi$ (see relations between $\xi$
and $n_s$, $r$ in
\cite{Anisimov:2008qs,Bezrukov:2008dt,Tsujikawa:2004my}).  In case it
is large, the laboratory detection will still be possible, while the
analysis of the current work should be rescaled for larger value of
the self-coupling constant $\beta$ and thus for stronger signal.

Note also that the inflaton model discussed in this paper has been
suggested in Ref.~\cite{Shaposhnikov:2006xi} as an extension of the
$\nu$MSM~\cite{Asaka:2005an,Asaka:2005pn}. In that joint model the
inflaton vacuum expectation value provides the three sterile neutrino
of $\nu$MSM with Majorana masses, while decays of inflatons in the
early Universe to the lightest sterile neutrino produce the dark
matter.  In Appendix \ref{sec:nuMSM} we check, that the limits
presented in $\nu$MSM itself do not limit the interval for the
inflaton mass obtained in this paper. On the contrary, the upper limit
on the inflaton mass constrains the mass of dark matter sterile
neutrino in this model. This proves that inflation can be directly
tested in fully realistic extensions of the SM.

We are indebted to M.~Shaposhnikov for valuable and inspirational
discussions throughout the work, P.\ Pakhlov for discussion on recent
Belle results, V.~Rubakov for valuable comments.  D.G. thanks EPFL for
hospitality.  The work of D.G. was supported in part by the Russian
Foundation for Basic Research (grants 08-02-00473a and 07-02-00820a),
by the grants of the President of the Russian Federation
MK-1957.2008.2, NS-1616.2008.2 (government contract 02.740.11.0244),
by FAE program (government contract $\Pi$520), by the Scopes grant of
Swiss National Science Foundation and by the grant of the Russian
Science Support Foundation.

%%%%%%%%%%%%%%%%%%%%%%%%%%%%%%%%%%%%%%%%%%%%%%%%%%%%%%%%%%%%%%%%%%%%%%%%
\appendix
\section{The $\nu$MSM extension}
\label{sec:nuMSM}

In Ref.~\cite{Shaposhnikov:2006xi} the inflaton model considered in
this paper has been implemented in the framework of
$\nu$MSM~\cite{Asaka:2005an,Asaka:2005pn}.

Neutrino oscillations is the only direct evidence of physics beyond
the Standard Model of particle physics.  Cosmology provides two other
evidences---dark matter and baryon asymmetry of the Universe---which
are not explained within SM if General Relativity is a correct theory
of gravity. These three problems can be addressed within a neutrino
minimal Standard Model ($\nu$MSM)~\cite{Asaka:2005an,Asaka:2005pn},
suggested as a minimal version of SM capable of explaining all of the
three. Enlarged by the additional scalar
field~\cite{Shaposhnikov:2006xi} coupled to the SM Higgs boson via
tree-level-scale-invariant interaction, $\nu$MSM provides both
early-time inflation and common source of electroweak symmetry
breaking and of sterile neutrino masses. This one-energy scale
model\footnote{For a discussion of gauge hierarchy problem in models
  of this type see, e.g.\ Ref.~\cite{Shaposhnikov:2007nj}.}  is a
minimal, full and self-consistent extension of the Standard Model of
particle physics.

This extension implies addition to the model \eqref{4*} three new
singlet right-handed neutrinos $N_I$, $I=1,2,3$ with the Lagrangian
\[
  \L_{\nu\mathrm{MSM}}
     =  i\bar N_I\gamma_\mu\d^\mu N_I 
        + \l F_{\alpha I}\bar L_\alpha N_I\tilde H 
        -\frac{f_I}{2}\bar N_I^c N_I X+\mathrm{h.c.}\r
  \;,
\]
where $L_\alpha$ ($\alpha=1,2,3$) are charged lepton doublets and
$\tilde H=\epsilon H^*$, where $\epsilon$ is $2\times2$ antisymmetric
matrix and $H$ is the Higgs doublet.

To begin with, let us discuss the allowed ranges of the Yukawa
coupling constants $f_I$ (Yukawas $F_{I\alpha}$ are irrelevant for
this study). The flatness of the inflaton potential implies smallness
of the quantum corrections to the quartic coupling
\eqref{quartic-from-CMB}. Requiring again smaller than 10\%
contribution we obtain
\begin{equation}\label{neutrino-yukawas-upper-limit}
  f_I < 1.5\times 10^{-3}
\end{equation}
and get for the sterile neutrino masses
\begin{equation}\label{neutrino-mass-upper-limit}
  M_I=f_I\, \la X\ra <
   270\cdot 
  \l \frac{m_\chi}{100\MeV} \r 
  \l \frac{1.5\times 10^{-13}}{\beta}\r^{1/2} 
  \l \frac{f_I}{1.5\times 10^{-3}}\r
  \GeV \;. 
\end{equation}
A lower limit on the masses of the two heavier sterile neutrinos,
$M_{2,3}$, follows from successfulness of the Big Bang Nucleosynthesis
(BBN), that does not constrain the inflaton mass.

The lightest sterile neutrino in the model \eqref{4*} can be stable at
cosmological time-scale and comprise the dark matter of the Universe.
In this case, the sterile neutrino is quite light, $M_1\lesssim
1$~GeV, and should not get thermalised in primordial plasma, to be
viable dark matter.  The latter is natural, since stability at
cosmological time-scale implies extremely small (if any) mixing with
active neutrino via $F_{1\alpha}$, and small mass implies tiny
coupling to inflaton. Light inflatons are in equilibrium in the
primordial plasma and decay to the lightest sterile neutrino mostly at
the temperature $T\sim m_\chi$. They provide contribution of the
sterile neutrino $N_1$ to the energy density of the Universe today
\cite{Shaposhnikov:2006xi}
\begin{equation}\label{neutrino-to-Omega}
  \Omega_N=\frac{1.6 f(m_\chi)}{S}\cdot \frac{\beta}{1.5\times10^{-13}}
  \cdot 
  \l  \frac{M_1}{10\keV}  \r^3 \cdot 
  \l  \frac{100\MeV}{m_\chi}  \r^3
  \;, 
\end{equation}
where $S>1$ is a dilution factor accounting for a possible entropy
production due to the late decays of the heavier sterile
neutrinos~\cite{Asaka:2006ek} and the function $f\l m_\chi\r $ is
determined by the number of degrees of freedom $g_*\l T\r$ in a
primordial plasma at inflaton decays.  It changes monotonically from
$0.9$ to $0.4$ for inflaton mass from 70~MeV to 500~MeV and for
heavier inflaton can be approximated as $f\l m_\chi\r\simeq \left[
  10.75/g_*\l m_\chi/3\r\right]^{3/2}$.  Other mechanisms can also
contribute to the dark matter production. Hence, at a given $m_\chi$
equation \eqref{neutrino-to-Omega} implies the \emph{upper limit}
\begin{equation}\label{upper-limit-on-DM-mass}
  \frac{M_1}{10\keV}<
  \l \frac{S}{6.4\, f(m_\chi)}\r^{1/3}
  \l \frac{\Omega_N}{0.25}\r^{1/3} 
  \l \frac{1.5\times10^{-13}}{\beta}\r^{1/3}
  \l  \frac{m_\chi}{100\MeV}  \r
  \;.
\end{equation}
The corresponding upper limit on $f_1$ supersedes
\eqref{neutrino-yukawas-upper-limit}.

There are several notes in order. First, from
\eqref{upper-limit-on-DM-mass} one concludes, that with the inflaton
mass in the range \eqref{light-inflaton-upper-limit},
\eqref{light-inflaton-lower-limit}, the lightest sterile neutrino can
\emph{naturally} be a Warm Dark Matter candidate. Second, with model
parameters tuned within their ranges to minimise r.h.s.\ of the
inequality \eqref{upper-limit-on-DM-mass}, one does not exceed the
lower bound on the sterile neutrino mass, $M_1\gtrsim 1.7$~keV, from
the study \cite{Gorbunov:2008ka,Boyarsky:2008ju} of the dark matter
phase space density in dwarf spheroidal galaxies.  Hence, in this
model limits from the dark matter sector do not shrink the allowed
parameter region, and in particular, the inflaton mass range. Third,
maximising r.h.s.\ of inequality \eqref{upper-limit-on-DM-mass} one
obtains an \emph{upper limit} on the lightest sterile neutrino mass,
\begin{equation}
\label{A5}
  M_1\lesssim 13 \cdot  \l  \frac{m_\chi}{300\MeV}  \r  
  \l \frac{S}{4}\r^{1/3} \cdot \l \frac{0.9}{f\l
  m_\chi\r}\r^{1/3}\keV
  \;.
\end{equation}
Hence, given the range \eqref{XXX} the lightest sterile neutrino is
significantly lighter than electron (and other charged SM
fermions). Since the light inflaton decays to SM fermions due to
mixing with the Higgs boson, these partial decay rates are
proportional to the squared masses of the corresponding fermions
similar to its decay rates to sterile neutrinos.  The lightness of the
dark matter sterile neutrino guarantees, that inflaton decays to the
dark matter neutrino never suppresses its decay branching ratios to
visible channels.

{}From the formulas above one concludes that the limits on the
inflaton mass obtained in this paper are not affected due to
additional constraints typical for $\nu$MSM. At the same time, the
limits \eqref{A5}, \eqref{XXX} imply that the dark matter is lighter
than about 100~keV and with account of the allowed Higgs boson mass
(see Sec.~\ref{sec:Model})\@.  Once the inflaton is found the upper
limit \eqref{A5} on the dark matter neutrino mass will be settled. And
vice versa, once the dark matter neutrinos are found, eq.~\eqref{A5}
fixes a \emph{lower} limit on the inflaton mass; this limit can
supersede limit \eqref{LLL} from direct searches, if
$M_1\gtrsim10$~keV.

Note that inflaton can decay into sterile neutrinos $N_{2,3}$ if
kinematically allowed.  The decay mode into the not-dark-matter
sterile neutrinos $N_{2,3}$ (invisible mode) has the width
\[
  \Gamma_{\chi \to N_I N_I} = \frac{\beta M_{I}^2}{8\pi m_\chi}
  \l 1-\frac{4m_f^2}{m_\chi^2} \r^{3/2}\;.
\]
This formula is very similar to (\ref{eq:gammall}).  This decay mode
can even be dominant, if the sterile neutrinos $N_{2,3}$ are the
heaviest fermions between the kinematically allowed ones.  As far as
$M_{2,3}\lesssim m_\pi$ is disfavoured by the BBN, this can be
relevant for inflaton masses above approximately 300~MeV.  As a
consequence, inflaton lifetime can be somewhat shortened as compared
to the results presented in Fig.~\ref{infdecays}b.

%%%%%%%%%%%%%%%%%%%%%%%%%%%%%%%%%%%%%%%%%%%%%%%%%%%%%%%%%%%%%%%%%%%%%%%%
\bibliographystyle{JCAP-hyper}
\bibliography{refs}

\end{document}